\documentclass[%
 reprint,
 amsmath,amssymb,
 aps,
 prl,
 twocolumn
]{revtex4-2}

\usepackage{graphicx}
\usepackage{dcolumn}
\usepackage{bm}
\usepackage{hyperref}
\usepackage{color}

\newcommand{\rzero}{r_0}

\newcommand{\vo}{v_o}
\newcommand{\rd}{r_d}

\newcommand{\cvo}{\tilde{v}_o}
\newcommand{\br}[1]{\left\langle#1\right\rangle}
\newcommand{\bigI}{\tilde I}

\begin{document}


\title{Innovation-exnovation dynamics on trees and trusses}

\author{Edward D.~Lee}%
 \email{edlee@csh.ac.at}
\author{Ernesto Ortega-D\'iaz}
\affiliation{%
 Complexity Science Hub, Metternichgasse 8, Vienna, Austria
}%

\date{\today}

\begin{abstract}
Innovation and its complement exnovation describe the progression of realized possibilities from the past to the future, and the process depends on the structure of the underlying graph. For example, the phylogenetic tree represents the unique path of mutations to a single species. To a technology, paths are manifold, like a ``truss.'' We solve for the phase diagram of a model, where a population innovates while outrunning exnovation. The dynamics progress on random graphs that capture the degree of historical contingency. Higher connectivity speeds innovation but also increases the risk of system collapse. We show how dynamics and structural connectivity conspire to unleash innovative diversity or to drive it extinct.
\end{abstract}


\maketitle


Innovation and its complement exnovation describe the constant churn of biological and social systems \cite{hochbergInnovationEmerging2017, leeIdeaEngines2024}. Evolutionary processes on one hand are driven by mutation and selection on top of competitive dynamics. A classic example is Van Valen's Red Queen, where competitors must constantly develop new strategies to keep up with the changing competitive landscape \cite{vanvalenNewEvolutionary1973}. In society, we often think of the Schumpeterian rise and fall of industries as a crucial element of the free market in the process of creative destruction \cite{schumpeterCapitalismSocialism1950}. Innovations are successful novelties \cite{erwinConceptualFramework2021} that arise from combinatorial search and randomness in the process of discovery \cite{younInventionCombinatorial2015}. One insightful formulation of potential novelties is the P\'{o}lya urn \cite{priceGeneralTheory1976, triaDynamicsCorrelated2015}, which has been extended to include variations such as social interactions that bias the search process \cite{iacopiniInteractingDiscovery2020}. Alternatively, innovation is formulated as the discovery of a percolating path \cite{silverbergPercolationModel2005}. To characterize innovativeness, many paradigms focus on scalar measures like changes in productivity \cite{aghionModelGrowth1990} or diversity \cite{triaDynamicsCorrelated2015}--even if the actual process is substantially more complex \cite{kavlakEvaluatingCauses2018}. On the other side is the process of exnovation, sometimes framed as a problem of extinction  \cite{newmanModelMass1997}, of forgetting \cite{amatoDynamicsNorm2018}, or of obsolescence \cite{priceGeneralTheory1976}, but which has generally received less attention. 
Thus, these processes constitute two complementary sides of change in living systems: innovation refers to the process via which biological or social systems realize new and impactful possibilities and exnovation how that set is truncated through obsolescence, desuetude, forgetting, and extinction. 

We propose a simple model to account for the race between innovation and exnovation. 
We model the dynamics of the ``space of the possible'' (SOP) \cite{kauffmanInvestigations2000}, the set of innovations often represented as technologies (e.g.~patents \cite{valverdeTopologyEvolution2007}), mutations (e.g.~genotypes, phenotypes \cite{erwinConceptualFramework2021}), or cultural practices (e.g.~spelling norms \cite{kolodnyGameChangingInnovations2016, amatoDynamicsNorm2018}). We formalize this as a directed graph $\mathcal{G}(\mathcal{X}, E)$ with vertices, or sites, $x\in\mathcal{X}$ representing possibilities and edges $E$ connecting innovations that lead from one to the other. The graph consists of $K$ parallel branches distinguishing phylogenetic taxa or as separate lines of inquiry. The space is bounded by Kauffman's ``adjacent possible'' (the set of all things one step away from what we know \cite{kauffmanInvestigations2000}) to the ``adjacent obsolescent'' (the set of all things one step away from exnovation \cite{leeIdeaEngines2024}). A natural and important extension is to hypergraphs \cite{iacopiniGroupInteractions2022}, but we do not consider these here. Thus, the space grows into the adjacent possible and shrinks away from the adjacent obsolescent in kind of competition between the two fronts.

\begin{figure}
	\centering
	\includegraphics[width=.8\linewidth]{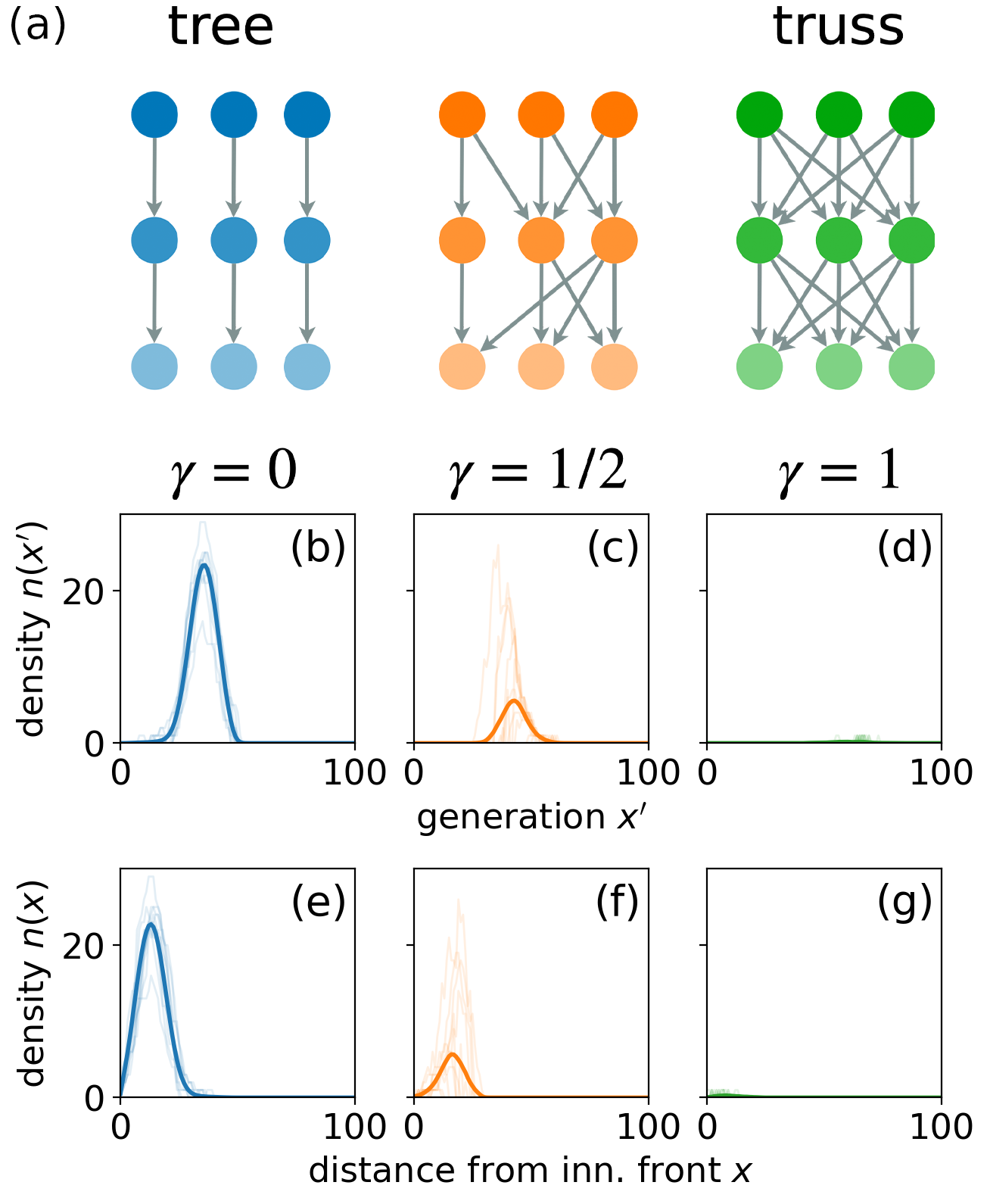}
	\caption{(a) Space of the possible, as a directed graph $\mathcal{G}$, extending into the adjacent possible at the bottom and disappearing at the adjacent obsolescent at the top. Connectivity $\gamma$ ranges from 0 (tree) to 1 (truss) to mimic phylogeny and technological history for branch number $K=3$. (b-d) Population density $n(x,t)$ for varying $\gamma$ in automaton simulation in unpinned, free coordinate system, or where $x'$ corresponds to the layer of the lattice. Thick line is average over $K=50$ branches and replicas $R=10^2$. Faint lines show a few individual replica branches. (e-g) Agent population in pinned coordinate system, where $x$ is the number of layers from the innovation front. The drop in density near the innovation front is the ``pseudogap.'' Averaged over $10^2$ replicas, $r_0=1/5$, $r_d=0.98$, $v_o=1/25$, $I=2$, $K=50$.}\label{fig:example}
\end{figure}

The classic representation of the SOP in biology is the phylogenetic tree of life, where every branching point indicates the emergence of a new species \cite{raboskyPhylogeneticTests2017}. This means that a species is the result of following a unique path of mutations through evolutionary history. On the other hand, we might think of technological development as less constrained to a fixed order. Imagine the alternative history of motor transportation, where the electric vehicle first becomes dominant rather than the internal combustion engine, the world where quantum mechanics comes after general relativity, or Cixin Liu's fictional, Trisolaran world where the widespread use of computers precedes that of the transistor. In this simplified framing, biological history is physically contingent, or a tree-like SOP (in the classical sense \footnote{This is a simplification of the modern understanding, which does not adhere strictly to the tree-like structure especially in the microbial world \cite{bushmanLateralDNA2002, daubinHorizontalGene2016}.}), whereas technological development is like a truss, where the next innovation can follow from any in the previous generation as we show in Figure~\ref{fig:example}a \footnote{This is of course a difference in definition of graph, where the former represents a physical description and the latter a functional one.}. To traverse the extremes, we take a single connectivity parameter $\gamma$ that ranges from $\gamma=0$ (a tree) to  $\gamma=1$ (truss) and corresponds to the probability that any site from the previous generation leads into an adjacent branch in the following generation in addition to its own branch. Thus, the connectivity allows us to explore a range of graph structures that titrate between the essence of a contingent versus a path-independent SOP.

Agents such as species or firms possess certain innovations, which we model as occupying sites in the SOP. As a general population dynamics intended to capture a wide range of scenarios, we consider several essential terms \cite{leeIdeaEngines2024}: we include the total influx of new agents as a rate $r_0$ (e.g.~reproduction or startup firms), agents leaving the system with rate $r_d$ (e.g.~death or bankruptcy), and mimetic innovation as rate $r$ as replication towards the direction of innovation (e.g.~mutation or spin-off firms). At the innovation front, agents cannot simply copy through mimetic innovation, but must also extract the site from the adjacent possible, the ease of which we capture with innovativeness factor $I$ to modulate replication at the front. Finally, we specify the exnovation velocity $v_o$ at which the adjacent obsolescent progresses and removes nodes, simultaneously excising the populating agents from system dynamics. All the rates can be scaled in terms of the replication rate $r$ without changing the following equations, leaving us with four parameters. Taking this general framework with a minimal parameterization of the population dynamics, we explore what happens as the structure of the SOP changes. 

The aforementioned elements correspond to the averaged population dynamics for $n(x, t)$, the number of agents occupying site $x$ at time $t$. After rescaling time by the replication rate $rt\rightarrow t$ and all other rates to be unitless, we obtain
\begin{align}
\begin{aligned}
    \partial_t n(x,t) = \frac{r_0}{|\mathcal{G}(t)|} - r_d n(x,t) + \sum_{x' \in \partial^+_x} \frac{n(x', t)}{|\partial^-_{x'}|}.
\end{aligned}\label{eq:mft1}
\end{align}
The SOP appears in Eq~\ref{eq:mft1} in terms of its size $|\mathcal{G}(t)|$, the number of sites between the adjacent possible and to the adjacent obsolescent. The second term captures agent death. Additionally, we denote the set of sites downstream of site $x$ as $\partial^{-}_x$, upstream as $\partial^+_x$, and their number with absolute values such that the last term captures how agents replicate uniformly across downstream sites. Eq~\ref{eq:mft1} captures agent population dynamics but not how the fronts move.

The innovation front is defined as the set of sites that are upstream of the adjacent possible $x\in\partial_{-1}^+$. The innovation front consists of sites in the SOP, and the adjacent possible is the set of sites that are one step away from being innovated into the graph \cite{kauffmanInvestigations2000}. Discovery of the next innovation is not the same as mimetic replication, so the innovativeness accounts for how easily $I>1$ (or laboriously $I<1$) they are found. The graph grows with the summed discovery rate at each site on the innovation front (as if each agent were independently attempting to push into the adjacent possible) and the typical number of downstream nodes that enter the SOP $\br{|\partial_{-1}^-|}$, or all together $I\br{|\partial_{-1}^-|} \sum_{x'\in \partial_{-1}^+} n(x',t)$ \footnote{Such a formulation also implies that the front spreads to all downstream nodes, which has the advantage that no trapped nodes are left behind fronts.}. As the complement, the adjacent obsolescent is the set of all points in the SOP that are one step away from exnovation. Falling behind the adjacent obsolescent excises the population from future dynamics, here captured as an abrupt loss in population, $n(x,t\geq t') = 0$ for all $x$ behind the front, $x\in\partial^+_\mathcal{O}$, for all time beyond the time of exnovation $t'$. Taking the two fronts and considering the case when the system size is large, the rate of change in the size of the SOP is 
\begin{align}
	\partial_t|\mathcal{G}(t)| = I\br{|\partial_{-1}^-|} \sum_{x\in \partial_{-1}^+} n(x,t) - v_o \br{|\partial_\mathcal{O}^-|}\sum_{x\in\partial^+_\mathcal{O}}1\label{eq:graph size mft}
\end{align}
Eq~\ref{eq:graph size mft} accounts for tension between a growing innovation front and elimination behind the exnovation front. In many systems, innovation and exnovation are coupled, but we take the simplifying assumption that they are parameterized by fixed rates and use those fixed rates as control parameters to study system behavior. The population dynamics in Eqs~\ref{eq:mft1} and \ref{eq:graph size mft} represent a distilled version of more complicated innovation-exnovation dynamics, but the nonlinearity in the rate of entering agents per site means that the dynamics, both stochastic or deterministic, are nontrivial.

To verify the calculations that we develop, we construct a highly parallelized, GPU-accelerated, and scalable automaton simulation, allowing us to access the full stochasticity of the system behind the mean-field equations. We use the numerical computation library JAX for random number generation and matrix multiplication to simulate the movement of the fronts and agent reproduction and death on a random graph. We incorporate an adaptive timescale that changes at each time step to ensure that the steps are sufficiently small to keep front velocities in the linear regime. Agent dynamics are Poisson distributed and are not sensitive to corrections from the linear approximation. Thus, we have a way of checking how the mean-field approximations describe a more complete instantiation of the dynamics.

Under weak coupling of the branches, the solution to Eqs~\ref{eq:mft1} and \ref{eq:graph size mft} leads to a rounded, Gaussian-like shape in the density function as in Figure~\ref{fig:example}a as the locations of the two fronts diffuse. The decay towards the leading edge of the branch reflects the tendency of the innovation front to deplete itself given that it tends to speed up when agents accumulate. The decay near the lagging edge represents the variability in the location of the exnovation front, behind which the density is zero. Most agents are typically found in between the two fronts for long-lived innovation trains, leading to a single-humped innovation train that drifts to the right.
	
This formulation, however, does not account for the fact that the density relative to the innovation front is of primary interest. We would like to incorporate this aspect into the mean-field equations. To track the innovation front, we pin each branch of the graph onto a unidimensional coordinate system such that the innovation front is fixed at $x=0$, the adjacent possible at $x=-1$, and a ``pseudogap'' in the density describes $x\geq0$. Then, we assume branch symmetry to take an average over the densities on each of $K$ respective branches. If the typical branch has time-dependent length $L(t)$, the exnovation front is located at $x=L(t)-1$. This is consistent with a  mean-field, continuous-time formulation for the population dynamics,
\begin{align}
    \partial_t n(x, t) &= \frac{\rzero}{L(t)} - \rd n(x, t)  + n(x+1, t) -\notag\\ 
    &\bigI\,  n(0, t)\Big [n(x, t) - n(x-1)\Big] \label{eq:mft3}\\
    \partial_t L &= \bigI \, n(0,t) - \cvo.\label{eq:mft4}
\end{align}
Eq~\ref{eq:mft3} accounts for per site growth, death, replication, and in the last term drift in the coordinate system due to the movement of the innovation front. Eq~\ref{eq:mft4} describes the rate at which the distance between the innovation and exnovation fronts changes (in the large $L$ limit). Both equations implicitly incorporate aspects of structural connectivity as represented by modified exnovation velocity $\cvo$ and modified innovativeness $\bigI$.

The modified terms represent faster front velocities because i) there are multiple paths through which to advance and ii) because fronts may be superseded when adjoining, further advanced branches feed into a given branch. Consider the fully connected limit $\gamma=1$, where the corrected velocity for either exnovation or innovation must be $\cvo = v_o (1+\gamma(K-1))$ as all leading fronts (on each branch) move together. This argument does not generalize to $\gamma<1$ because the leading fronts are not synchronized. 

Take the front that is furthest ahead to be at $y=0$, i.e., no other branch is further ahead. There are two scenarios for a branch with front at $y\geq0$:  it moves with probability $v_odt$ or it does not move with probability $1-v_odt$ in short time $dt\ll1$. In the former case, the front moves by one lattice site, and it expands out to a fraction $\gamma$ of its neighbors. The effective velocity at which the front on a single branch moves is then $v_o(K^{-1}+\gamma(1-K^{-1}))$. There are $K q(y)$, for cumulative distribution $q(y)\equiv\sum_{y'=0}^y p(y')$, exnovation fronts in any layer $y$ (the cumulative term accounts for the assumption that any exnovated site can propagate downstream). Now, we average over all the layers $v_o\sum_{y=0}^\infty p(y)(1+\gamma(K-1)q(y)) = v_o(1+\bar{q}\gamma(K-1))$, taking $\bar{q}=\sum_y p(y)q(y)$. The innovation front is different because only every site can only be innovated once, and so the leading site on a given branch is the only one that can advance into the adjacent possible. As a result, at each layer $y$ only a fraction $p(y)K$ can propagate the front, or a velocity $n(0)I (1+\bar{p}\gamma(K-1))$, given $\bar{p} \equiv \sum_{y=0}^\infty p(y)^2$ and the averaged, stationary innovation front density $n(0)$. 

Consider now the second contribution to the velocity. The front does not move, but it effectively jumps ahead to $y'-1$ if an adjacent branch at $y'<y$ feeds in ahead of it. For the exnovation front, each layer ahead of it has $q(y')K$ fronts that individually move with rate $v_o$ and feed into the branch at layer $y$ with probability $\gamma$, or $v_o \gamma K \Delta y_1$ and $\Delta y_1\equiv \sum_{y=1}^\infty p(y) \sum_{y'<y}q(y')(y-y'+1)$. The argument is the same for the innovation, except that each layer ahead $y'<y$ has $p(y')K$ fronts to obtain $\Delta y_2\equiv \sum_{y=1}^\infty p(y) \sum_{y'<y}p(y')(y-y'+1)$.

\begin{figure}
\centering
	\includegraphics[width=\linewidth]{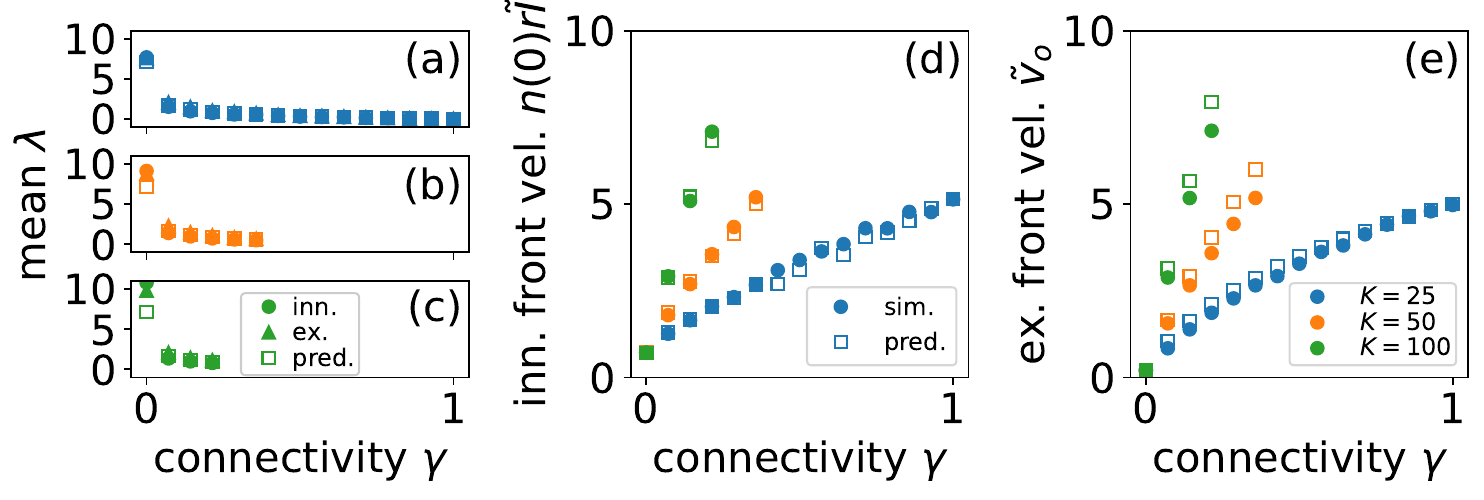}
	\caption{(a-c) Predicted average distance to the leading front $\lambda$ against automaton simulation for innovation and exnovation fronts given branch numbers $K=25$, $K=50$, and $K=10^2$, respectively. For sufficiently large $K$ and $\gamma$, the system collapses and no value is given. Comparison of front velocities for (d) innovation and (e) exnovation fronts. Averaged over $R=25$ replicas, $r_0=1/5$, $r_d=0.98$, $v_o=1/25$, $I=2$.}\label{fig:mft fronts}
\end{figure}

To complete the calculation, we must find $p(y)$. In the limit $\gamma\rightarrow1$, where fronts are thinly distributed around the leading front, they escape promotion to the front only rarely. This picture suggests as an approximation the Poisson distribution $p(y)= e^{-\lambda}\lambda^{-y}/y!$ (here we are considering either many random replicas or $K\gg1$). The mean $\lambda$ is the mean distance $\bar{y}$. Note that this form for the probability distribution has the correct limits $\lim_{\gamma\rightarrow0} p(y) = \delta(y)$ and a uniform distribution when $\gamma\rightarrow1$. By writing down the relationship between leading sites that move between $y=0$ and $y=1$, we obtain a self-consistency condition under steady state $\gamma = e^{-\lambda}/(1+\lambda)$, implying an implicit solution for $\lambda$ in terms of $\gamma$.

All together, the corrected front velocities when $K\gg1$ are given by
\begin{align}
\begin{aligned}
	\cvo&\equiv\vo[1+\bar{q}\gamma K + \gamma K\Delta y_1],\\
	\bigI&\equiv I[1+\bar{p}\gamma K + \gamma K\Delta y_2].
\end{aligned}\label{eq:big I}
\end{align}
The first two terms in each line of Eq~\ref{eq:big I} account for self-promotion of a site and the second term promotion from branches that are further ahead (note that this approximation ignores the fronts that are temporarily behind others on the same branch). Eq~\ref{eq:big I} obeys the expected limits: $\lim_{\gamma\rightarrow0} \cvo=\vo$ and $\lim_{\gamma\rightarrow1} \cvo=v_o(1+\gamma K)$ as well as $\lim_{\gamma\rightarrow0} \bigI=I$ and $\lim_{\gamma\rightarrow1} \bigI = I(1+\gamma K)$, or the cases when the branches are fully independent or perfectly synchronized.

We show how the calculations for the innovation and exnovation front velocities compare with the automaton simulation in Figure~\ref{fig:mft fronts}. The two agree well over most of the range for three different example branch numbers $K=25$, $K=50$, and $K=10^2$. The average distance $\lambda$ of fronts from the leading front disagree the most as $\gamma\rightarrow0$, but this has little effect on the averaged front velocities because jumps become rare and negligible in this limit. This is clear in panels d and e, where we compare the predicted and simulated front velocities. For small $\gamma$, the analytical approximation is close to simulation. For larger $\gamma$, we find stronger disagreement because the connectivity is weak enough that $p(y)$ has a wide distribution, jumps are large, and this exaggerates the error from the Poisson approximation to $p(y)$. As $\gamma\rightarrow1$, we expect such errors to diminish because the distribution $p(y)$ is narrow and self-promotion becomes the dominant contributor to the modified velocities. Along with the modified velocities, Eqs~\ref{eq:mft3} and \ref{eq:mft4} describe the averaged dynamics given distance $x$ to the innovation front and over the ensemble of random graphs that are characterized by $\gamma$. For both fronts, redundant paths to children nodes increases velocity over independent branches, distinguishing the truss from the tree. 

Na\"{i}vely, we expect three types of solutions \cite{leeIdeaEngines2024}: (i) \textit{runaway} where the innovation front outpaces exnovation; (ii) \textit{collapse} when the exnovation front catches up with innovation; and (iii) \textit{balanced} where innovation and exnovation velocities match. From \textit{in silico} experiments, we observe more exotic behaviors such as multiple fixed points, depending on initial conditions. We also observe a separation of timescales, where the total density settles rapidly at a fixed value but the system size $L(t)$ relaxes much more slowly. We also find the trivial fixed point $L=0$ can have a large stochastic basin of attraction. To better characterize the observations, we propose an analytic approximation to Eqs~\ref{eq:mft3} and \ref{eq:mft4}.

\begin{figure}
\centering
	\includegraphics[width=\linewidth]{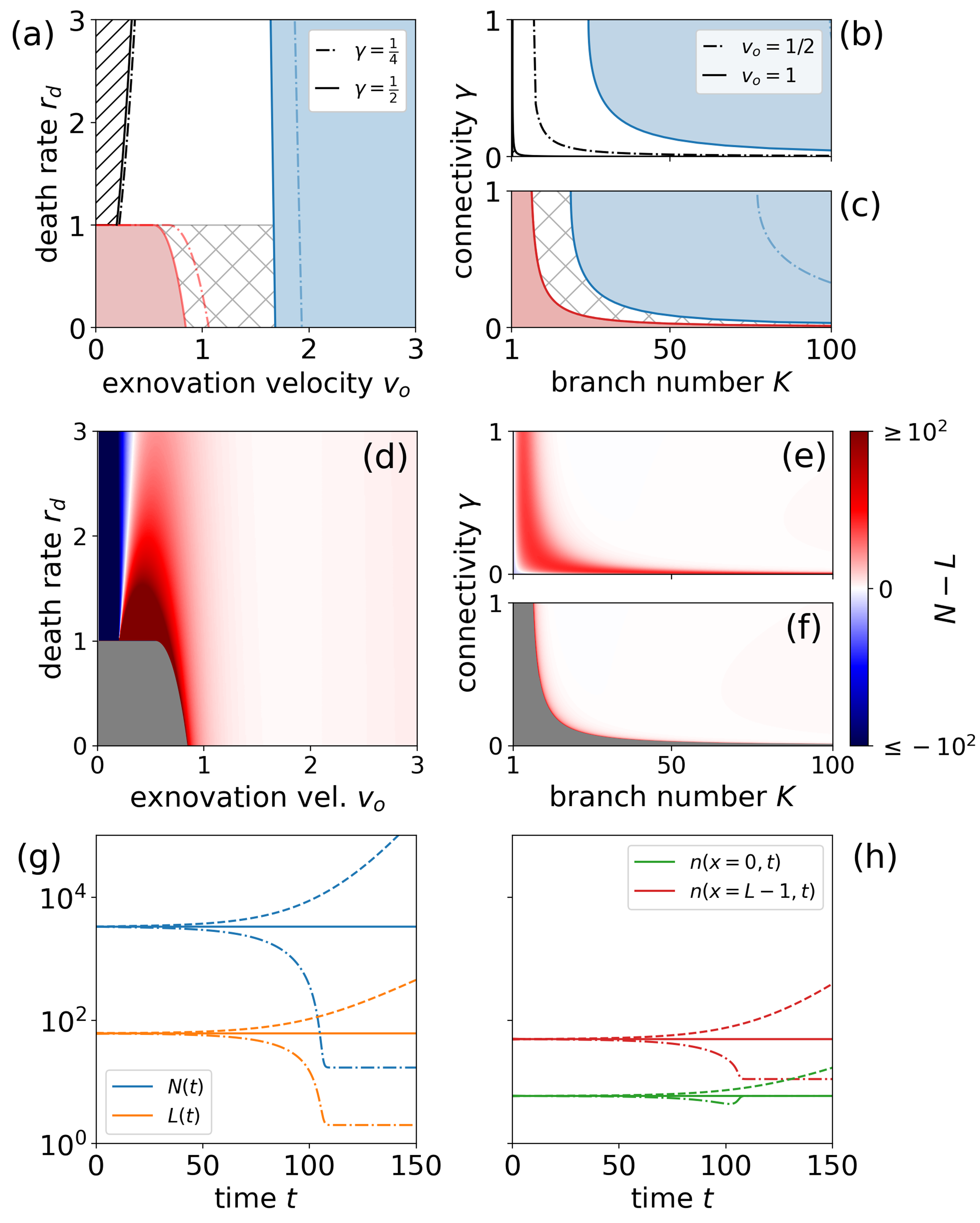}
	\caption{(a) Dynamical phase diagram by death rate $\rd$ and exnovation velocity $\vo$ (other parameters $r_0=125$, $I=2$, $K=12$). (b, c) Structural phase diagram by connectivity $\gamma$ and branch number $K$ ($r_0=25$, $I=2$, $r_d=5/2$ for top panel and $r_0=50/3$, $I=2$, $r_d=5/3$ for bottom panel). Lines distinguish values of connectivity $\gamma$ or exnovation velocity $v_o$. Red regions are runaway, blue collapsed, white steady state, single-line hatched low-density, and cross-hatched bistable phases from compartment model. (d-f) Comparison of total agent number and lattice length, $N_+-L_+$, with parameters from panels a, b, and c, respectively. Runaway region in gray. Bifurcating trajectories on initial conditions by (g) total lattice number $N(t)$ in blue and length $L(t)$ in orange and (h) number at innovation front $n(0,t)$ in green and exnovation front $n(L-1,t)$ in red. Each line represents a different initial condition perturbed from the unstable solution from Eq~\ref{eq:compartment ss}, where for dashed $L(0)=L_-+1$, solid $L(0)=L_-$, and dot-dashed $L(0)=L_--1$. The latter converges to the stable solution $N_+$ and $L_+$.}\label{fig:phase space}
\end{figure}

The fronts constitute key elements of the model, and they constitute ``interfaces'' that are coupled through the flow of agents in the ``bulk'' in between. This is the compartment model. The bulk has length $L^*(t) = L(t) - 2$, accounting for the two fronts, and number of agents $N^*(t) = N(t) - n(0, t) - n(L-1, t)$, discounting the front densities. To couple the bulk with the fronts, we make the approximation that the density inside the bulk can be treated homogeneously, or that $n(x,t)=N^*(t)/L^*(t)$ for $1\leq x \leq L-2$. This approximation obtains a closed set of four equations,
\begin{align}
\begin{aligned}
	\partial_t N &= \rzero - \rd N + N-n_0 - \cvo n_{L-1}\\
	\partial_t L &= \bigI n(0) - \cvo\\
	\partial_t n_0 &= \rzero/L - \rd n_0 + \\
		&\quad(N-n_0-n_{L-1})/(L-2) - \bigI n_0^2\\
	\partial_t n_{L-1} &= \rzero/L - \rd n_{L-1} - \cvo(n_{L-1}-\\
		&\quad(N-n_0-n_{L-1})/(L-2)).
\end{aligned}\label{eq:compartment}
\end{align}
Eq~\ref{eq:compartment} can be solved at stationarity to obtain the order parameters
\begin{align}
\begin{aligned}
	N_{\pm} &= \left(B\pm \sqrt{C}\right)/2A \\
	L_{\pm} &= \left(B'\pm \sqrt{C'}\right)/2A'.
\end{aligned}\label{eq:compartment ss}
\end{align}
Additionally, we find a quadratic solution for $n_{L-1}$, and $n_0=\cvo/\bigI$. We do not enumerate the terms in $A$, $B$, $C$, and their prime counterparts here. Eq~\ref{eq:compartment ss} represents two solutions, where the radicals for both $N$ and $L$ share the same sign, but the negative solution only holds for \mbox{$r_d<1$}. Non-physical (negative or complex) solutions reveal when the system displays infinite growth or collapse, and we solve for these lines to draw the phase diagram in Figure~\ref{fig:phase space}a-c.

Collapse implies the disappearance of the bulk, or $L\leq2$. From the bulk-interface approximation, we get as the condition delineating the collapsed regime,
\begin{equation}
    r_d = \left(-b+\sqrt{b^2-4ac}\right)/2a \label{eq:collapse regime}
\end{equation}
with $a=2\cvo(1+\cvo)$, $b=4\cvo^3-\bigI \rzero(1+\cvo)$, and $c=2\cvo^3(\cvo-1)-\bigI \rzero(1+\cvo^2)$, having disposed of the three other non-physical solutions. Eq~\ref{eq:collapse regime} is displayed as the blue region in Figure~\ref{fig:phase space}a, where the solid blue line delineates this regime as where the exnovation velocity exceeds innovation's. When we increase $\gamma$ from $\gamma=1/2$ to $\gamma=1$ (dashed blue line), the collapsed region grows. In panels b and c corresponding to $r_d>1$ and $r_d<1$, respectively, we show again the boundary of the collapsed regime but for fixed dynamical parameters and as we vary structural parameters. The collapsed regime emerges for sufficiently large branching number and connectivity---this signals the role of branch-spanning connections in facilitating system collapse.

Divergence, in contrast, is when either $N\rightarrow\infty$, which implies that $L\rightarrow\infty$, or just $L\rightarrow\infty$ regardless of $N$. Thus, we can search for where $L$ diverges. At those lines, $L$ crosses over to negative or imaginary solutions, so we solve for where $\Im[L]\geq0$. In the dynamical phase diagram as in Figure~\ref{fig:phase space}a, the physical solution (out of four) delineating the regime of complex $L$ is
\begin{align}
	\rd &= \frac{1}{4\cvo}\left[\bigI \rzero + 3\cvo - \cvo^2 - \left(\left(\bigI \rzero + 3\cvo - \cvo^2\right)^2+\right.\right.\notag\\
		& \left.\left. 8\cvo\left(\cvo^2 + \cvo^3 - 2\sqrt{\bigI \cvo \rzero (1 + \cvo^2)}\right)\right)^{1/2}\right].\label{eq:dynamical runaway}
\end{align}
This alone, however, is insufficient because $L$ becomes negative but remains real on a line that cuts through the runaway region in the bottom left of Figure~\ref{fig:phase space}a. The crossover to negative values is at $\rd=1$ and for all values $\vo$ smaller than the value at which Eq~\ref{eq:dynamical runaway} peaks, $\vo=\vo^*$, and bends down and away from $\rd=1$. These two conditions together delineate the red, runaway regime. As we decrease connectivity $\gamma$ from 1/2 to 1/4, the size of the runaway region expands, again showing that less connectivity enhances system stability.

To the right of the runaway region, we find a bistable regime indicated by the cross-hatched region in Figure~\ref{fig:phase space}a. Here, both solutions from Eq~\ref{eq:compartment ss} are viable (in particular $N_-$ and $L_-$ satisfy the steady state conditions only in this regime), but the $+$ solutions are stable and the $-$ solutions unstable. Fluctuations around the unstable solution to lead to divergence or to the stable solution. We show example trajectories of this outcome in Figures~\ref{fig:phase space}f-g. 

In the remaining region, we find a single stable and finite solution in the white regions of Figure~\ref{fig:phase space}a-c. Of particular curiosity the emergence of a stable low-density phase, which is where the density drops below unity, $\mathbb{R}[N-L]<0$. In panel a, this is where $\vo$ is small but $\rd>1$ as indicated by the hashed region. This is the blue region in Figure~\ref{fig:phase space}d. This is a ``Byzantine'' phase of high but mostly static diversity: innovation is not particularly fast, but exnovation is slow enough to allow the system to be large. Moving over to the structural phase space in panels b and c, we see that at small $K$ that connectivity has little effect on density. In this regime, $K$ is sufficiently small that connectivity no longer moves us between phases, a kind of insensitivity to interdependence. Finally, we note that appreciable densities occupy relatively small volumes of the stable phases such as the thin sliver bordering runaway for $r_d<1$ in panel f. This is especially notable as $K$ becomes large, where the dense, stable phase represents a tiny sliver of parameter space. This indicates that stability is in this sense unusual in large systems and that thriving population dynamics are limited to a small region of that space.

\begin{figure}\centering
\includegraphics[width=\linewidth]{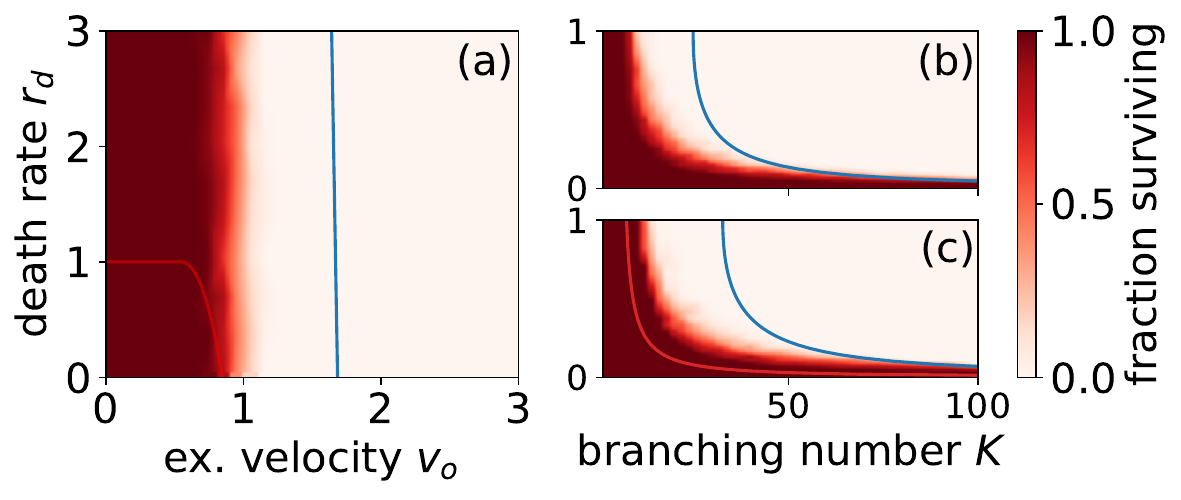}
\caption{Fraction of surviving branches calculated over $R=50$ replicas at time $t=20$, given steady-state initial conditions from stable compartment model solutions corresponding to $N_-$ and $L_-$ (except at runaway zones, where initial conditions are capped at $N\leq10^4$ and $L\leq10^2$). Red and blue contours delineate runaway and collapsed regions from Figure~\ref{fig:phase space}a-c. (a) Dynamical phase diagram. (b) Structural phase diagram $r_d>1$ and (c) $r_d<1$. Simulation parameters from Figure~\ref{fig:phase space}a-c.}\label{fig:automaton}
\end{figure}

To check the stability of the mean-field solutions under stochastic dynamics, we simulate them with the automaton. We initialize random replicas with conditions as given by the compartment model at steady state, taking uniform density in the bulk $n(x,t=0)=N_+^*/L_+^*$ and the steady-state interface values and evaluate the fraction of surviving branches per replica at time $t=20$ across the phase diagrams as shown in Figure~\ref{fig:automaton}---since collapse is a fixed point, the fraction cannot increase with time. As expected, empty branches are more likely as we approach the boundaries of the collapsed phase in blue, where small system size makes it more likely that random fluctuations leave the system in the collapsed configuration. The compartment model formulation captures the general contours of the automaton simulation, although the trivial attractor is larger than as indicated by the approximation. This reflects the error from our approximation of $\cvo$ in Figure~\ref{fig:automaton}a and the large $K$ approximation in panels b and c. For large $K$ in the structure space (panels b and c), the transition from collapse to runaway is compressed into a small and diminishing range of $\gamma$ as we increase $K$. Since $\gamma$ is symmetric with respect to the number of incoming upstream edges and outgoing downstream edges, the compression indicates the importance of outgoing edges to the dynamics, which determines how quickly density drops at the innovation front.

We present and solve a simplified model that explicitly accounts for the innovation and exnovation fronts (e.g., Schumpeterian creative destruction \cite{schumpeterCapitalismSocialism1950}) and the population dynamics of agents living on the changing graph. Crucially, we account for divergent (hierarchical) and convergent (truss-like) paths of development that distinguishing biology-like vs.~technology-like systems \cite{kuhnEssentialTension2000, soleEvolutionaryEcology2013}. Combining a compartment model with numerical calculation, we reveal the multiple phases of such dynamics and how the boundaries between them change with the number of branches and the connectivity between them. Thus, we identify how dynamics and structure of the SOP conspire to either enhance or hamper the race between innovation and exnovation.

A particular question on which the model touches is how few systems remain constant in structure over time. The ``tree of life'' is rife with cross-cutting connections as organisms exchange DNA or RNA \cite{maynardsmithOriginsLife2000, daubinHorizontalGene2016}. Species experience population bottlenecks or separate as ecosystems divide as with the disappearance of Beringian Strait or as beautifully demonstrated by Darwin's finches. Economies may decouple, scientific fields split, or new interdisciplinary centers arise. 
Here, we explore a simple model that displays the consequences of varying branch interaction in terms of the number of parallel branches and the degree of connectivity between them. Perhaps counterintuitive to the notion that crossing bridges helps to develop new ideas, we find that increasing connectivity between adjacent branches also risks system instability, even if it increases the innovation rate. This trade-off between system survival and speedy innovation suggests a type of evolutionary or cultural speed limit \cite{zeldovichProteinStability2007} that takes into account the underlying graph structure. As a second observation, we find that the regions of parameter space that display a thriving population dynamics, where more than one agent can be found per innovation, constitutes a relatively small fraction of the space, centered around SOPs with fewer branches and weaker connectivity. This is perhaps relevant to the emergence of modularity in adaptive, resource-constrained systems.


\begin{acknowledgments}
We thank Rudi Hanel and Vito Servedio for discussion. EDL acknowledges funding from the Austrian Science Fund grant ESP-127. EO acknowledges funding from the Austrian Federal Ministry for Climate Action, Environment, Energy, Mobility, Innovation and Technology (2021-0.664.668) and the City of Vienna.

\end{acknowledgments}

\bibliographystyle{unsrt}
\bibliography{biblio,refs} 
\end{document}